# How a Klein-Nishina Modified Eddington limited accretion explains rapid black hole growth in the early universe

Jackson Frangos[1], Erick Rosen[2], Michael Williams[2], Chandra B. Singh[3], David Garofalo[2]

[1]Magnet Program, Joseph Wheeler High School, Marrietta, GA 30068, USA; jackson.frangos@wheelermagnet.com
[2]Department of Physics, Kennesaw State University, Marietta, GA 30060, USA; dgarofal@kennesaw.edu
[3]South-Western Institute for Astronomy Research (SWIFAR), Yunnan University, University Town, Chenggong, Kunming 650500, People's Republic of China; chandrasingh@ynu.edu.cn

**Abstract**

The discovery of quasars and their supermassive black holes (SMBHs) over $10^9$ M$_\odot$ merely hundreds of millions of years after the Big Bang generates tension with the idea of Eddington-limited accretion and pressures the community into exploring the concept of massive black hole seeds and/or super-Eddington accretion. The observation that many black holes have reached supermassive status while obeying the Eddington limit is puzzling as accretion models are not spherically symmetric. We address this issue by illustrating the physics behind a picture of inner disk accretion involving a geometrically thick, hot quasi-spherical flow and argue that such an inner region provides the radiation that instantiates the Eddington limit. Given the energetics of the inner disk edge, we show how the characteristic electron cross-section drops below its Thomson value, allowing black holes to grow rapidly despite being Eddington-limited. Indeed, after implementing a modified cross-section calculated via the Klein-Nishina Formula, we find that SMBH formation time drops by up to 47%. In this context, we show how a $10^9$ M$_\odot$ black hole can form from a seed 10 M$_\odot$ black hole within 500 Myr by way of accretion and mergers. While our picture is over-simplified and contrived in a number of ways that we discuss, we suggest that our scenario is interesting in that it offers a solution to two issues at the intersection of astrophysics and cosmology, namely the reason the Eddington limit is obeyed and how some black holes have grown rapidly despite that limit.

Unified Astronomy Thesaurus concepts: Active galactic nuclei (16); Quasars (1319); Supermassive black holes (1663); Galaxy evolution (594); Early universe (435); Active galaxies (17)

# 1. Introduction

Recent observational evidence obtained using instruments onboard the James Webb Space Telescope (JWST) has revealed a significant number of supermassive black holes (SMBHs) in the early universe (Bosman et al. 2024; Suh et al. 2024; Juodžbalis et al. 2024 and references therein). Such massive black holes (BHs) have also been identified by powerful jets observed with ALMA (Endsley et al 2023). Various models have been proposed to explain this rapid buildup to SMBHs, including relatively massive BH seeds accreting in the sub-Eddington regime or from light stellar remnants or heavy seeds with short phases of super-Eddington accretion rates ( e.g., Schneifer et al. 2023, Trinca et al. 2022). Other models include the evolution of radiatively inefficient accretion flow onto slowly rotating BHs (Inayoshi & Ichikawa 2024) or even radical ideas involving new cosmological models such as the $R_h = ct$ universe (Melia 2023, 2024 and references therein).

In this work, we revisit the issue of rapid SMBHs formation in the early universe using a novel picture for spherical accretion near the horizon of a BH that is motivated by analytic and numerical work. To understand the impact of this picture, we first develop a simple analytical model of SMBH formation based on a well-understood accretion scenario. We then implement the modified electron cross-section prescribed by our picture of inner accretion, which results in a marked decrease in SMBH formation time.

We start with a $10 M_\odot$ BH and grow it by alternating mergers and accretion. For simplicity, and to develop our algorithms, we initially assume the accretion events share the same timescale, which makes our mathematics simpler. We estimate the relaxation time following a merger which is the time prior to the onset of accretion, and the accretion time, and sum them to estimate the total time for SMBH buildup in terms of the number of mergers. We then make our exploration more physically realistic by having the accretion time scale with the size of the system, connect the disk efficiency to the time of accretion due to black hole spin-up, and finally explore the physics of a new picture for accretion focused on the region near the inner disk edge where radiation pressure dominates the dynamics. While the scenario is made increasingly more physically well-motivated in Section 2, most of the astrophysics discussed there, hinges on well-understood mechanisms. It is not until we revisit the growth timescale assuming a spherical inner disk region in section 2d, that a new picture emerges. If our idea for the character of accretion in the inner region is reasonable, we show that it solves two issues. First, it motivates the unexplained observations that black holes tend to obey the Eddington limit (e.g. Kollmeier et al. 2006), and second, it explains how they can nonetheless grow rapidly. The conclusion is that a reasonable space exists within standard accretion models to grow SMBHs to $10^9$ $M_\odot$ prior to redshift 7.

## 2. Growth of black holes by mergers and accretion

a. Eddington-limited growth with equal accretion times

The Eddington accretion rate prescribes the accretion rate to be proportional to the BH mass via

$$dm_{BH}/dt = km_{BH}, \quad (1)$$

which can be solved to get the mass growth as

$$m_f = m_i e^{kt} \quad (2)$$

which describes the final mass of a BH of initial mass $m_i$ after accreting for some time t. The constant k is given by

$$k = \frac{4\pi G m_p}{\eta c \sigma_t} \quad (3)$$

Where G is the gravitational constant with a numerical value of 6.674 x $10^{-11}$ Nm$^2$/kg$^2$, $m_p$ is the mass of the proton which is 1.67 x $10^{-27}$ kg, c is the speed of light with its typical value of 299,792,458 m/s, $\sigma_t$ is the Thomson scattering cross-section of the electron which is 6.65 x $10^{-29}$ m$^2$, and η is the radiative efficiency of a thin accretion disk, generally taken as 0.1. We consider a BH of mass $m_i$ that undergoes n accretion events, wherein each accretion event is followed by a doubling merger. Normally, a merger results in some of the input mass being lost due to gravitational waves, leading the final black hole to have slightly less mass than expected. That being said, the goal of this paper is to develop a relatively simplistic model to establish a baseline for the feasibility of Eddington limited accretion. As such we are solely looking at the most rapid growth channel allowable, meaning we can reasonably assume the target black hole tends to merge with BH's slightly larger than it, leading to a mass doubling merger. The idea behind the doubling of the mass is that we are looking for the most rapid growth channel. We therefore consider that the most massive BHs merge. If we assume the BH merges with a significantly more massive BH than itself, or with one that is less massive, that implies that we are following the growth of a BH that is not the most rapid. While such mergers are the majority, they are not the most rapid. That being said, a brief analysis of SMBH formation with non-doubling mergers is included in appendix A.

The algorithm that captures the growth cycle described above is

$$m_f = \underbrace{\overbrace{m_i \cdot 2e^{kt_1}}^{m_i\ for\ t_2} \cdot 2e^{kt_2}}_{m_i\ for\ t_2} \cdot 2e^{kt_3} \ldots 2e^{kt_n} \quad . \qquad (4)$$

Equivalently,

$$m_f = m_i \cdot 2^n \prod_{i=1}^{n} e^{kt_i} \qquad (5)$$

We also consider the relaxation time, or the time it takes the system to begin accreting following the merger. The timescale ansatz, for simplicity, scales linearly with the black hole mass, capturing the idea that smaller systems will form their accretion disks more quickly. Given that two BHs of $10^8$ M$_\odot$ undergo a 75 Myr relaxation time after merging (Somerville et al. 2001), the total relaxation time after a merger between two BHs of mass $m_{premerger}$ is given by

$$t_{relax} = m_{premerger} \cdot \frac{75}{10^8}. \qquad (6)$$

The total time the BH spends growing to $10^9$ M$_\odot$, will, therefore be expressed as

$$t_{tot} = t_{accrete} + t_{relax} \qquad (7)$$

where $t_{accrete}$ is the sum of the individual accretion times

$$t_{accrete} = \sum_{i=1}^{n} t_i \quad,$$

and $t_{relax}$ is the sum of the individual relaxation times

$$t_{relax} = \sum_{i=1}^{n} t_{ri} \quad.$$

For the nth relaxation time, $t_{rn}$, the premerger black hole mass is equal to $m_f(n)$ divided by two, as the black hole has yet to undergo the latest merger. Symbolically we have

$$t_{rn}(n) = m_i \frac{75}{10^8} 2^{n-1} \prod_{i=1}^{n} e^{kt_i}.$$

(8)

When added together to get the total growth time of a BH we have

$$t_{tot} = \sum_{i=1}^{n} t_i + m_i \frac{75}{10^8} \sum_{i=1}^{n-1} 2^{i-1} \prod_{j=1}^{i} e^{kt_j}.$$

(9)

For simplicity, we begin by assuming all accretion times $t_i$ to be equal. This is unrealistic, but allows us to produce simple analytic solutions that we can use to understand the physics as we add complexity. This assumption yields

$$t_1 = \frac{\ln\left(\frac{m_f}{m_i 2^n}\right)}{kn}$$

(10)

which when implemented into the general growth expression results in a final expression

$$t_{tot}(n) = \frac{\ln\left(\frac{m_f}{(m_i) 2^n}\right)}{k(3.1536 \cdot 10^{13})} + \frac{75}{2 \cdot 10^7} \left( \left(\frac{m_f}{m_i}\right)^{\frac{1}{n}} \frac{\frac{m_f}{m_i} - 1}{\left(\frac{m_f}{m_i}\right)^{\frac{1}{n}} - 1} \right)$$

(11)

which obtains $t_{tot}$ in millions of years as a function of the number of mergers needed to grow the BH. Letting $m_i = 10$ $M_\odot$ and $m_f = 10^9$ $M_\odot$ we plot $t_{tot}$ vs n, the number of mergers, shown in Figure 1. Graphs of both $t_{accrete}$ vs n and $t_{relax}$ vs n, are shown in Figures 2 and 3.

Note that in Figure 1, adding additional mergers exhibits a diminishing returns effect, each subsequent added merger lowers the total time less effectively. Conceptually, this is because mergers grow the BH quickly compared to accretion. Figure 2 shows a negative linear relation between time spent accreting and the number of mergers, while also showing an exponential relation between total merger time and number of mergers.

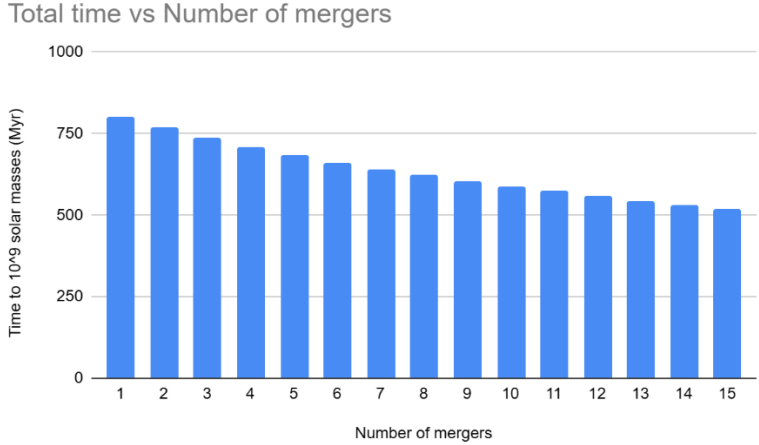

Figure 1: Final growth time as a function of the number of mergers, for equal accretion times

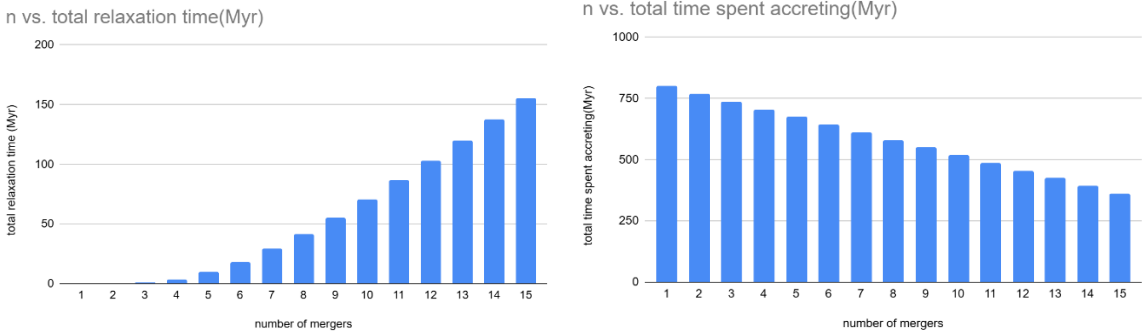

Figure 2: Total relaxation time (left) and total accretion time (right) as a function of merger number n.

b. Eddington limited growth with non-equal accretion times

Our next step is to make the accretion timescales more realistic. Since bigger BHs will be embedded in larger protogalaxies and eventually larger galaxies, accretion will last longer for successive accretion events. We assume that every accretion event lasts a time that is some constant multiple, λ, of the previous accretion time. While it is reasonable to assume that larger systems lead to longer-lasting accretion events, the timescale for successive accretion events is not expected to follow a precise algorithm. We motivate this algorithm because it produces a simple analytic expression that allows for ease of calculation and such that it is compatible with analysis of clustering measurements suggesting that when the BH is massive enough to be a quasar, it accretes for tens of millions of years so that it is constrained to accretion episodes on the order 100 million years (Martini & Weinberg 2001; White et al 2012; Conroy & White 2013; La Plante & Trac 2016). We will see that in the final analysis, our fiducial path ends with a 200 million solar mass BH experiencing a final accretion event lasting 134 million years. Thus, for the i$^{th}$ accretion time, we have

$$t_i = \lambda^{i-1} t_0$$

(12)

where $t_0$ is the time of the initial accretion event following merger 1. Substituting this into equation 2 and solving for $t_0$ yields

$$t_0 = \frac{(\lambda - 1) \ln\left(\frac{m_f}{m_i 2^n}\right)}{k(\lambda^n - 1)}$$

(13)

which leads to our final expression for growth time with non-constant accretion times as

$$t_{tot} = \frac{\ln\left(\frac{m_f}{m_i 2^n}\right)}{k(3.1536 \times 10^{13})} + m_i \frac{75}{10^8} \sum_{i=1}^{n-1} \left(2^{i-1} \prod_{j=1}^{i} \left(e^{k \lambda^{i-1} t_0}\right)\right).$$

(14)

Using equation (14), we make new figures for relaxation time, accretion time, and total time to grow the BH. As before, we assume an initial BH mass of $m_i = 10$ M$_\odot$ and an ending BH mass of $m_f = 10^9$ M$_\odot$. A well-motivated estimation of λ is outside the scope and goal of this paper, as it likely depends on a number of different conditions relating to the environment of the early universe. That being the case, it is worth noting that a higher value of λ leads to a higher concentration of rapid, low-mass black hole mergers that dominate the early growth phase of our black hole. This early concentration of rapidly merging low-mass BHs aligns to some degree with ideas in the community about hierarchical or runaway mergers, so a higher value of λ is not entirely unmotivated. However, the idea that mergers are waiting to happen following prescribed, constant ratio accretion times is one of the more contrived features of our model and we will discuss these limits in more depth in the conclusion. Using a range of λ values such that 1 < λ < 2, we produce Figures 3 and 4. These values imply that we allow the next accretion event to last at most twice as long as the previous one. Any values significantly greater than two would result in unrealistically short times between early mergers (i.e. $t_0$ in equation (13) goes like the inverse of λ to the n-1) and would not be compatible with observations from clustering of quasars as discussed above for the late mergers. Perhaps the best argument for that range of λ values comes from a physical argument. Mergers of larger systems funnel more gas into the nucleus, but larger systems are more complex, and gas ends up being distributed in random ways that are not compatible with direct fueling of the BH. Hence, we think of λ=2 as an upper limit but surely larger than 1. For those reasons, we arrive at 1 < λ < 2.

Beyond this, black hole duty cycle analysis on the Illustris simulation (DeGraf & Sijacki 2017) find that Black hole duty cycles scale with $M_{bh}^{0.3-0.5}$ which suggests a similar scaling for accretion event times. Such a scaling combined with doubling mergers results in a lambda value of between 1.3-1.5. While duty cycle scaling's are not exact analogs for accretion event scaling, they do provide some additional grounding for our final growth channel identified in section 2d in which we use λ = 1.5 to achieve SMBH formation in less than 500 Myr.

Unlike in iteration one, the diminishing returns behavior of additional mergers quickly disappears as λ increases in value, shown in Figure 3, causing the graph of $t_{tot}$ to approach a linear curve. This can be understood to result from higher λ values being associated with a larger percentage of mergers occurring earlier. This concentration of mergers early on makes the average relaxation time very small. As λ grows larger, the average relaxation time approaches zero, making it so that every merger is equally efficient, leading to a linear curve. As shown in Figure 4, different λ values do not affect the relationship between accretion time and number of mergers.

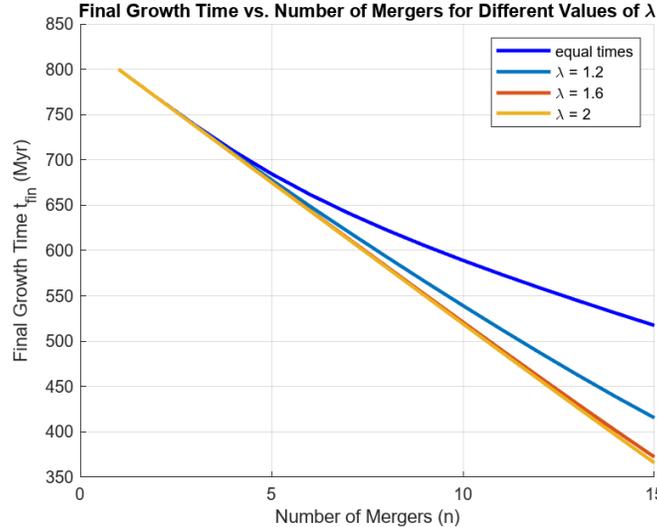

Figure 3: Final growth time curves with respect to allowed mergers for variable accretion times.

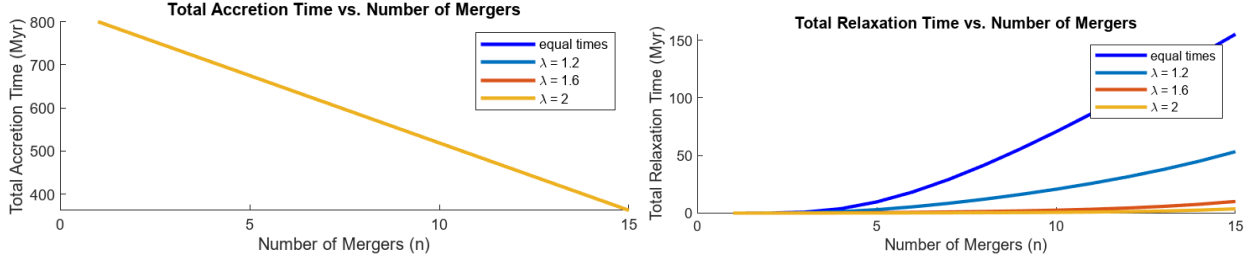

Figure 4: Total accretion time (left) and total relaxation time (right) for variable accretion times.

c. Eddington-limited growth with varying accretion efficiency

In this subsection, we explore a more realistic accretion efficiency by recognizing that the efficiency depends on black hole spin. For accretion triggered by mergers onto a newly merged BH that spins rapidly but such that the accretion settles into counter-rotation, it will take about 100 Myr for the black hole to spin up to high spin in corotation where the efficiency is largest (Kim et al 2016 section 3.1). Most of that time is associated with corotation between BH and accretion disk since a counter-rotating high-spinning BH spins down to zero in $8 \times 10^6$ yrs at the Eddington limit (Kim et al 2016), i.e. independent of BH mass as shown in Appendix B. The standard efficiency of 0.1 is augmented at high corotation to a value of 0.4. We don't track black hole spin in any detailed way (for such an exploration see Kim et al 2016). Instead, we assume that accretion events that last 100 Myr or more can be treated as having an average efficiency of 0.4. Meanwhile any accretion events lasting under 100 Myr are treated with the standard 0.1 efficiency.

The equation for BH mass becomes

$$m_f = m_i \cdot 2^n \cdot \prod_{i=1}^{j} e^{k \lambda^{i-1} t_0} \cdot \prod_{l=j+1}^{n} e^{\frac{k}{4} \lambda^{l-1} t_0}$$

(15)

where the first product term represents the growth of the BH during all accretion events under 100 Myr while the second product term represents the growth from all accretion events over 100 Myr. in time. While n still represents number of mergers, j is a new term that represents the number of accretion events that are under 100 Myr long. It can be calculated via the ceil function ⌈ ⌉,

$$j = \left\lceil \log_\lambda \left( \frac{100}{t_0} \right) \right\rceil.$$

(16)

The equation for $m_f$ was converted into a finite geometric series and subsequently simplified to achieve the following expression for $m_f$.

$$m_f = m_i \cdot 2^n \cdot \exp\left(t_0 k \left(\frac{\lambda^j - 1}{\lambda - 1} + \frac{1}{4} \cdot \frac{\lambda^n - \lambda^j}{\lambda - 1}\right)\right).$$
(17)

As this equation does not have an analytic solution for $t_0$ in terms of n we turn to a numerical solution for growth time calculations. Recall from iteration one that $t_{relax}$ depends on $m_f$, as such the final expression for $t_{tot}$ is

$$t_{tot} = \frac{1}{3.1536 \times 10^{13}} \sum_{i=1}^{n} \lambda^{i-1} t_0 + \sum_{i=1}^{n-1} \frac{m_f(i)}{2}.$$
(18)

$M_f$ was substituted to obtain

$$t_{tot} = \frac{1}{3.1536 \times 10^{13}} \sum_{i=1}^{n} \lambda^{i-1} t_0 + \sum_{i=1}^{n-1} m_i 2^{i-1} \exp\left(t_0 k \left(\frac{\lambda^j - 1}{\lambda - 1} + \frac{1}{4} \cdot \frac{\lambda^i - \lambda^j}{\lambda - 1}\right)\right).$$
(19)

As in previous iterations, we make plots of total, accretion, and relaxation, times. These timescales are shown in Figures 5 and 6 and color-coded to different values of λ. The addition of a more physical variable accretion efficiency caused a sharp jump in growth timescales when compared to previous iterations, placing almost all new growth timescales outside of the 800 Myr range set by JWST observations. Additionally, it is worth noting that while in the previous iterations a higher λ value led to smaller growth times, it now appears that lower λ values are most optimal. Lower λ values tend to generate fewer accretion events over 100 Myr timescales, which leads to generally decreased radiative efficiencies and hence faster accretion. This implies that rapid BH buildup in the early universe occurred in mergers that did not result in the most efficient funneling of gas into the nuclear region. This we believe to be a new and interesting result worth further exploration.

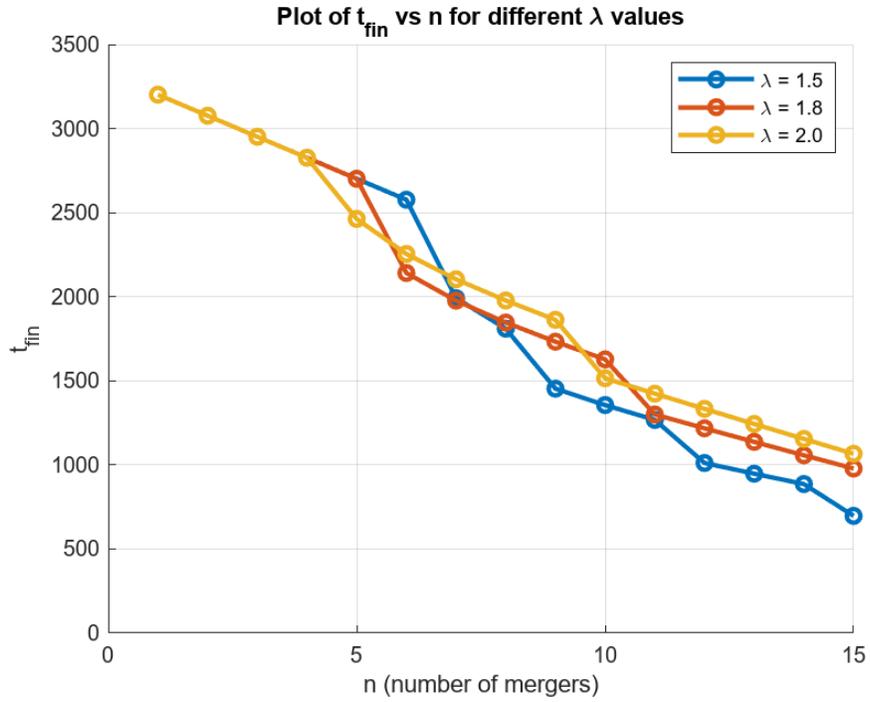

Figure 5: Final growth time curves with respect to allowed mergers, variable accretion time and radiative efficiency. Times given in Myr

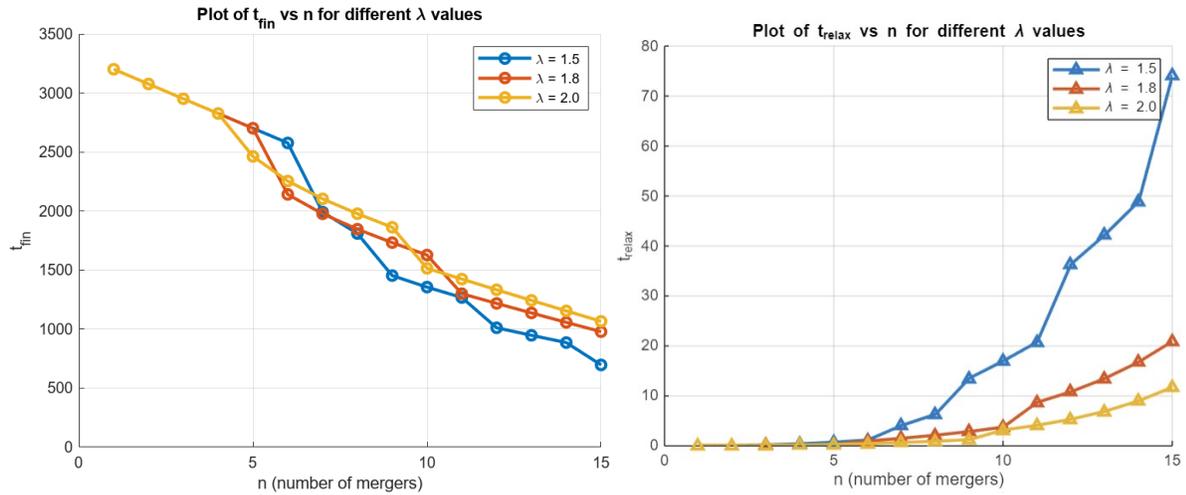

Figure 6: Total accretion time (left) and total relaxation time (right) as a function of n, for both variable accretion times and variable radiative efficiency. Times given in Myr.

d. Eddington-limited growth with relativistic electron cross-section

So far, our exploration of black hole accretion has operated under a standard assumption that the radiation from the accretion disk has an equatorial component that instantiates the Eddington limit. This assumption rests on a weak foundation, since most of the radiation is expected to occur in a direction that is perpendicular to the accretion plane and because the Eddington limit assumes a spherical distribution of radiative matter. While we appeal to decades of simulation work, we exercise some creative license in postulating the existence of a quasi-spherical accretion flow in the inner region of the accretion disk. The inner disk taps into the deepest gravitational potential and is where the dissipation function reaches its peak (for analytic studies see e.g. Shakura & Sunyaev 1973). Numerical simulations (e.g. Liska et al. 2022) have shown that such conditions lead to a radiation-dominated region that generates a thick, or quasi-spherical, hot pressure-supported flow due to large-scale magnetic fields and radiative processes. This is the only region of the accretion flow that radiates quasi-spherically and we imagine that it is responsible for ensuring that the Eddington limit is obeyed. Because of the proximity of this region to the event horizon, the generated radiative photons have the highest energy and as such they determine a modified electron cross section which we calculate by using the Klein-Nishina expression (Ghisellini 2013) as follows.

$$\sigma = 2\pi r_e^2 \left[ \frac{1+\epsilon}{\epsilon^3} \left( \frac{2\epsilon(1+\epsilon)}{1+2\epsilon} - \ln(1+2\epsilon) \right) + \frac{\ln(1+2\epsilon)}{2\epsilon} - \frac{1+3\epsilon}{(1+2\epsilon)^2} \right] \quad (20)$$

with $\epsilon = E_e/M_e c^2$, the ratio of the photon energy to the electron rest mass energy.

Thus instead of the standard Eddington limit constant, k, we now have

$$k = \frac{4\pi G m_p}{\eta \sigma(\epsilon) c} \quad (21)$$

where $\sigma$ now acts as a function of $\epsilon$.

To understand the details of such hot inner accretion flow, we appeal to both analytic work (Meier 2002) as well as state-of-the-art two-temperature general relativistic magnetohydrodynamic simulations that have explored the plasma microphysics (Chael et al 2018). From a temperature in this inner hot flow of at least $10^9$ K (but possibly as high as $10^{11}$ K; see Wienke 1985), we obtain $\epsilon = 0.17$ as a typical value (e.g. Chael et al. 2018) by assuming that photon temperature is the same as the electron temperature close to the Eddington accretion rate. This is physically feasible in such a regime as the photons and ionized plasma components, namely electrons, are likely to undergo frequent energy exchange. The possible coexistence of such hot flows with cold, thin-disk accretion, is interesting to our study since it provides a natural explanation for both Eddington-limited accretion and for rapid growth due to the large $\varepsilon$ value. Figure 7 displays a schematic of this accretion structure. But such hot flows are essentially sandwiched between the inner edge of the thin disk and the BH horizon, which implies that the BH spin is not close to unity since a corotating accretion disk around a 0.998 spin BH has the disk inner edge at the event horizon and would therefore not provide any space for such hot flow to exist. As accretion spins the BH up to maximal value, the radiation pressure-dominated inner region would disappear, and the system would then experience a bout of super-Eddington accretion. There are, however, possible restrictions on the value of black hole spin that may ensure that the inner edge does not reach the event horizon, possibly due to magnetically driven winds that extract angular momentum from the disk (Blandford & Payne 1982) and/or from the black hole (Benson & Babul 2009).

Note from Table 1 that these high spin values tend to occur only for the last handful of accretion events. The highest value of $\varepsilon$ occurs nearest to the BH (see also Satapathy, Psaltis & Özel 2023 where the ion-to-electron temperature in the inner hot accretion flow was explored for a wide range of physical scenarios accounting for plasma properties, mass accretion rates, and BH spins). Semi-analytic calculations suggest these results are weakly dependent on BH mass and spin (see Figure 2 in Manmoto 2000 for spin dependence; Figures 7 and 10 in Manmoto, Mineshige & Kusunose 1997 for mass dependence).

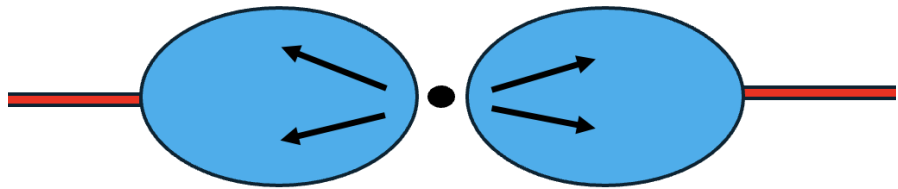

Figure 7: The hot inner region (blue) of a cold thin accretion disk (red) around a black hole (black circle) is supported by radiation pressure (arrows) that explains the Eddington limit in accreting black holes of all scales.

Using equation (20), we redo our usual calculations, first by evaluating a new cross-section for $\varepsilon = 0.17$ which lowers $\sigma$ from it's typical value of 6.65 x 10$^{-29}$ to a new value of 5.046 x 10$^{-29}$, thereby increasing the value of k. This means that a higher accretion rate is compatible with the Eddington limit. The results appear in Figure 8. With a $\lambda$ value of 1.5, 14 mergers, and a total growth time of 500 Myr, we have shown that such SMBHs can grow rapidly within the context of reasonable astrophysical parameters. Note that even though smaller merger values result in SMBH formation times that exceed those observed by the JWST, they are all markedly shorter formation times than those predicted by the model when using the standard Thomson cross-section. In some cases, this decrease in formation time can reach up to 47%, with an average decrease across all merger values of 32.1%. Thus, regardless of the feasibility of a 14 merger SMBH formation, we have still shown that accounting for a modified electron cross-section is critical in explaining rapid SMBH formation.

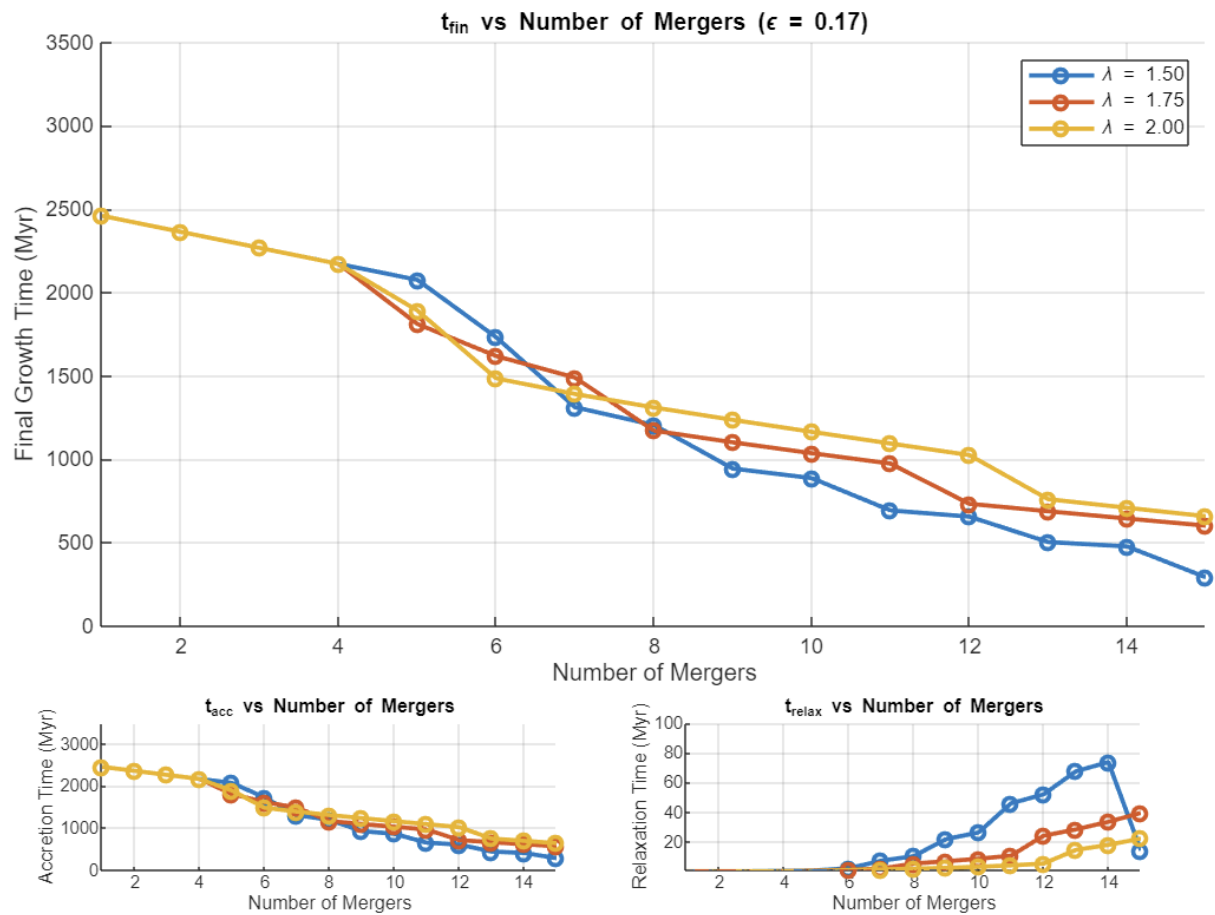

Figure 8: Total time (above), accretion time (lower left), and relaxation time (lower right) for Eddington-limited thin disks whose inner edge puffs up with cross sections determined by $\varepsilon = 0.17$.

A more detailed numerical analysis of the 14-merger growth channel allows us to look at the BH mass at each growth stage. The results, shown in Table 1, indicate that the BH only breaks one million solar masses during its last two mergers after only 178 Myr. From this it can be inferred that most of the mergers take place in small, rapidly merging protogalaxies. Our analysis has so far ignored the question of when our 10 $M_\odot$ BH was seeded. It is assumed that it was the product of the end state of a massive population III star of a few tens of solar masses, that was born circa 200 Myr after the Big Bang during Cosmic Dawn and lived about 10 Myr. This means that our $1.89 \times 10^8$ $M_\odot$ BH formed 270 Myr plus ~ 200 Myr which equals 470 Myr after the Big Bang resulting in the formation of $10^9$ $M_\odot$ SMBH ~670 Myr after the Big Bang.

| Merger number | Post-merger mass ($M_\odot$) | Total growth time (Myr) | Total accreting time (Myr) | Total relaxation time (Myr) |
|---|---|---|---|---|
| 1 | 20.4 | 0.689 | 0.689 | 0 |
| 2 | 42.1 | 1.72 | 1.72 | $7.65 \times 10^{-6}$ |
| 3 | 88.0 | 3.27 | 3.27 | $2.34 \times 10^{-5}$ |
| 4 | 188 | 5.60 | 5.60 | $5.64 \times 10^{-5}$ |
| 5 | 417 | 9.09 | 9.09 | 0.000127 |
| 6 | 970 | 14.3 | 14.3 | 0.000283 |
| 7 | $2.44 \times 10^3$ | 22.2 | 22.2 | 0.000757 |
| 8 | $6.86 \times 10^3$ | 33.9 | 33.9 | 0.00156 |
| 9 | $2.29 \times 10^4$ | 51.6 | 51.6 | 0.0041 |
| 10 | $9.87 \times 10^4$ | 78.1 | 78.1 | 0.0127 |
| 11 | $6.26 \times 10^5$ | 118 | 118 | 0.0497 |
| 12 | $7.06 \times 10^6$ | 178 | 177 | 0.284 |
| 13 | $1.89 \times 10^8$ | 270 | 267 | 2.93 |
| 14 | $1.00 \times 10^9$ | 474 | 401 | 73.8 |

Table 1: The first column gives the number of mergers the black hole has undergone, the second column is the total mass of the black hole directly after the indicated number of mergers, the third column is the total time the black hole took to get to that mass, and the fourth and fifth columns breakdown that total time into time spent accreting and time spent in post-merger relaxation periods, respectively.

3. Summary and conclusions

Of the millions of SMBHs that have been discovered across the Universe, the majority of them are observed at redshift below 7, which gives them enough time to grow to their measured masses within the confines of well-understood astrophysical processes such as Eddington-limited accretion. However, a minority of SMBHs have reached large masses and are observed at redshifts of 7 or above, producing tension with the conventional growth channel, which has led to the proliferation of ideas for rapid growth; many of which invoke super-Eddington accretion, massive BH seeds, or some combination of the two. We revisit the issue as a way of homing in on the exact nature of the tension by embracing an iterative approach designed to isolate and analyze the effects of varying basic astrophysical parameters.

We begin by making simple assumptions that allow us to determine the appropriate algorithm for obtaining the accretion time, relaxation time, and the total growth time for a 10 $M_\odot$ BH to grow to $10^9$ $M_\odot$. Our second iteration makes the accretion times more physical by scaling the time with the size of the system. If the system doubles in size after a merger, we assume the amount of fuel that ends up in the nucleus of the new protogalaxy/galaxy will grow by a similar factor. Next, we analyze the effects of changing radiative efficiency as governed by the BH spin. Over long periods of continuous accretion BHs tend towards highly prograde disk configurations, resulting in efficiencies of roughly 0.42. We therefore track efficiency by classifying accretion periods as either short accretion events (<100 Myr) or long accretion events (>100 Myr). Any accretion events that have lasted more than 100 Myr, even when starting from a retrograde configuration, will have had enough time to re-configure towards a high-prograde system. As such, all long events are treated with an efficiency of 0.40, while all short events are treated with an efficiency of 0.1. Up to this point, all of our work has been to establish a base analytical model derived from standard accretion theory. We then test this model both with and without the modified electron cross section prescribed by our picture of the hot inner flow. In doing this we find that accounting for the modified cross-section significantly lowers SMBH formation time.

It is important to acknowledge the fact that this study suffers from an admittedly contrived and somewhat "unphysical" analysis. We track black hole efficiency and spin in a simplified manner, optimistically model mass losses in mergers, and use linearly scaling relaxation timescales. Despite these limitations, we argue for and motivate the idea that early SMBH black hole formation can be explained via Eddington-limited accretion with the help of non-trivial cross sections motivated by the quasi-spherical flow between the ISCO and the event horizon. In doing so, we also offer a conceptual explanation as to why black holes, despite their apparent lack of radiative spherical symmetry, still obey the Eddington limit. Additionally, we have conservatively estimated inner flow temperatures at $10^9$ K, it is entirely possible that such flows could reach temperatures of $10^{11}$ K, which could lead to even more rapid accretion and quicker SMBH formation while still residing within the Eddington regime.

In conclusion, we have found a tentative rapid growth pathway to SMBH formation in the Eddington regime via the introduction of a modified electron cross-section. The ideal pathway involves 14 doubling black hole mergers with a concentration of low-mass rapid mergers in the early growth stages. Only two mergers occur between black holes of over $10^6$ $M_\odot$, and only one accretion event was longer than 100 Myr. Additionally, we propose that, supported by more in-depth numerical studies, an inner hot accretion flow sandwiched between the ISCO and the event horizon could both sufficiently explain the apparent adherence of growing BHs to the Eddington limit while also providing a growth pathway for early SMBH formation. Thus, beyond simply proposing a tentative growth pathway to SMBH formation, we have introduced and quantified the importance of a previously largely ignored factor in Eddington Limited accretion: the modified electron cross section described by the Klein-Nishina formula. Implementing this modification, regardless of factors like the number of mergers, pre-merger post-merger mass ratios, and individual accretion event length, tends to drive down SMBH formation times by around 30%. Thus, we can conclude that the modified electron cross section is an important term in explaining early SMBH formation within the bounds of Eddington-limited accretion and has the potential to eliminate the need for other more extreme alternatives, such as super-Eddington accretion.

Acknowledgments
CBS is supported by the National Natural Science Foundation of China under grant No. 12073021.

## Appendix A: Non-doubling Mergers

Thus far we have only explored the fastest possible reasonable growth channels for SMBH formation, hence we assumed doubling mergers. However, a significant number of BH mergers are non-doubling in nature. In order to validate our study's conclusion that accounting for a modified cross-section is important in rapid SMBH formation, we have also evaluated the modified cross-section's importance in non-doubling merger growth scenarios. In order to do so we replace equation 17 with the following more generalized version:

$$m_f = m_i \cdot r^n \cdot \exp\left(t_0 k \left(\frac{\lambda^j - 1}{\lambda - 1} + \frac{1}{4} \cdot \frac{\lambda^n - \lambda^j}{\lambda - 1}\right)\right). \quad (21)$$

Where r represents the average pre-merger to post-merger mass ratio. From this we derive equation 22, a modified version of equation 19 that similarly accounts for a general merger ratio:

$$t_{tot} = \frac{1}{3.1536 \times 10^{13}} \sum_{i=1}^{n} \lambda^{i-1} t_0 + \sum_{i=1}^{n-1} m_i r^{i-1} \exp\left(t_0 k \left(\frac{\lambda^j - 1}{\lambda - 1} + \frac{1}{4} \cdot \frac{\lambda^i - \lambda^j}{\lambda - 1}\right)\right). \quad (22)$$

Figure 9 shows the results of plotting equation 22 with r values of 1.3, 1.5, and 1.7. Plotting the results both with and without accounting for a modified cross-section clearly shows that, regardless of the merger ratio, accounting for the modified cross-section results in an average decrease in SMBH formation time by about 30%

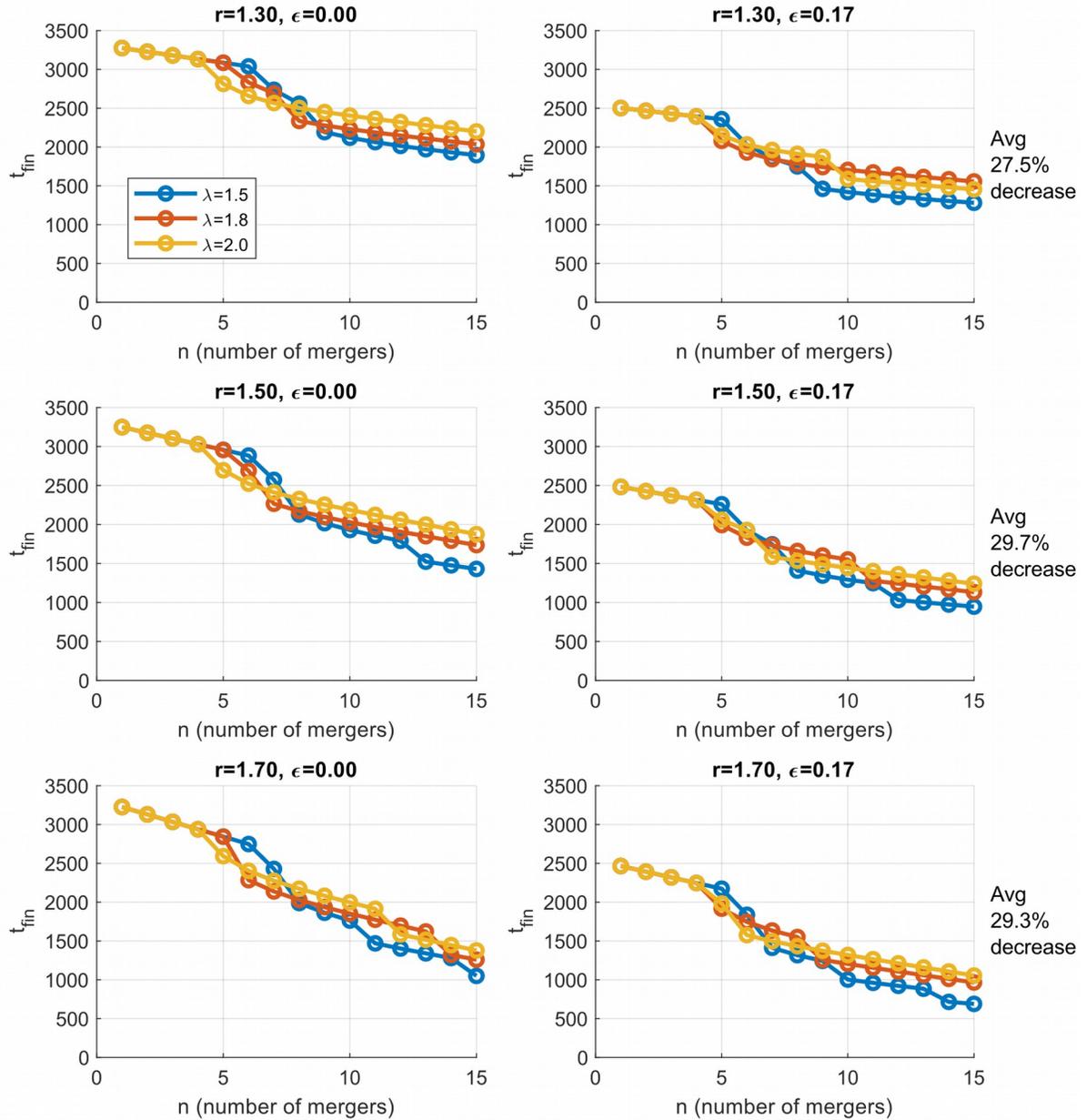

Figure 9 – Shows the total time to SMBH formation both without a modified cross-section(left) and with a modified cross-section(right). Each row shows the results for a different average post-merger pre-merger mass ratio, r, along with the average per cent decrease in SMBH formation time that occurs once we account for the modified electron cross section at each r value. Each color corresponds to a different ratio ($\lambda$) between subsequent accretion event lengths.

## Appendix B: The independence of Eddington-limited accretion time on BH mass

Here we show that the rate at which the dimensionless black hole spin changes with time, d$a$/dt does not depend on BH mass. The dimensionless black hole spin is related to the angular momentum $L$ of the black hole via

$$a = cL/(GM_{BH}^2)$$
(1)

with G Newton's constant, c the speed of light and $M_{BH}$ the black hole mass. We write this in terms of the infinitesimal increment in angular momentum delivered to the black hole by accreted gas as

$$dL = GM_{BH}^2 da/c.$$
(2)

A parcel of gas dm that is moving on circular orbits with speed v that reaches the disk inner edge at r, carries an infinitesimal amount of angular momentum. Thus, in magnitude

$$dL = dm\, v\, r.$$
(3)

Our goal is to reframe this in terms of the dimensionless black hole spin. This gives

$$GM_{BH}^2 da/c = dm\, v\, r.$$
(4)

On circular orbits and using Newtonian theory, we have

$$v = (GM_{BH}/r)^{0.5}.$$
(5)

The value of r at the inner edge varies as a function of black hole spin but is proportional to the BH mass. For zero spin, r = $6GM_{BH}/c^2$. For maximal spin this value drops to about 1.2 $GM_{BH}/c^2$. Given that we don't need the numerical coefficient for our purposes, we show the result for zero BH spin to get

$$GM_{BH}^2 da/c = dm\, (GM_{BH}/r)^{0.5} 6GM_{BH}/c^2 = dm(6)^{0.5}\, GM_{BH}/c$$
(6)

which then leads to,

$$da/dt = M_{BH}^{-1}(6)^{0.5} \, dm/dt.$$
(7)

To proceed we must obtain dm/dt which is the Eddington limit given by the balance of radiation pressure and gravity via

$$\eta c^2 dm/dt = 4\pi c G m_p M_{BH}/\sigma$$
(8)

with $\eta$ the radiative efficiency of the disk, c the speed of light, $m_p$ the mass of the proton, and $\sigma$ the electron cross section. From this balance we obtain the accretion rate as

$$dm/dt = 4\pi G m_p M_{BH}/(\eta \sigma c)$$
(9)

which replaces dm/dt in equation (7) to obtain

$$da/dt = (6)^{0.5} 4\pi G m_p/(\eta \sigma c).$$
(10)

The time to spin the BH up or down to any value at the Eddington-limit, therefore, is independent of BH mass.